\documentclass[twocolumn,aps,floats,showpacs,prb]{revtex4}
\usepackage{amsmath}
\usepackage{epsfig}
\usepackage{graphicx}
\usepackage{bm}

\def\gapp{\lower.35em\hbox{$\stackrel{\textstyle>}{\sim}$}}
\def\lapp{\lower.35em\hbox{$\stackrel{\textstyle<}{\sim}$}}

\begin{document}
\title{Fast and slow edges in bilayer graphene nanoribbons:\\ Tuning the transition from band- to Mott-insulator}

\author{Alberto Cortijo, L\'aszl\'o Oroszl\'any, and Henning Schomerus}
\affiliation{Department of Physics, Lancaster University,
Lancaster, LA1 4YB, United Kingdom}

\date{\today}
\begin{abstract}
We show that gated bilayer graphene zigzag ribbons possess a fast and a slow
edge, characterized by edge state velocities that differ due to
non-negligible next-nearest-neighbor hopping elements. By applying
bosonization and renormalization group methods, we find that the slow edge
can acquire a sizable interaction-induced gap, which is tunable via an
external gate voltage $V_{g}$. In contrast to the gate-induced gap in the bulk,
the interaction-induced gap depends non-monotonously on the on-site potential $V$.
\end{abstract}

%

\maketitle

\section{Introduction}

One of the most attractive properties of bilayer graphene  (which
is made of two coupled atomic layers of carbon) is the ability to
induce a tunable spectral gap $\Delta_V$ by applying a
perpendicular electric field.
\cite{TaisukeOhta08182006,McCann06,McCann06a,Castro07,Min08}
Rapid advances in patterning graphene on the nanoscale now make it
feasible to fabricate graphene ribbons with well defined edge
termination, \cite{Jiao09,Campos09} while experiments on narrow
ribbons show that they can display a gap $\Delta_W$ due to
transverse size quantization. \cite{Han07} The ensuing facility to
confine electrons in a controllable way via gate potentials and
patterning makes bilayer nanoribbons a promising candidate for
nanoelectronic applications. Systems suitable of being used as a
basis for a transistor should exhibit a high resistance in the off
state. In the case of nanoribbons with zigzag termination,
however, this requirement poses a problem, since the edges support
current-carrying states with energies inside the bulk gap, and a
much reduced hybridization gap $\Delta_h\ll\Delta_V,\Delta_W$ due
to exponentially small tunneling between the edges
.\cite{Castro08,Rhim08,Sahu08,Lima09} On the other hand, several
novel concepts (notably, valleytronics \cite{Rycerz07}) exploit
the existence and specific properties of such edge states.

\begin{figure}
\raisebox{.8cm}{\includegraphics[width=.8\columnwidth]{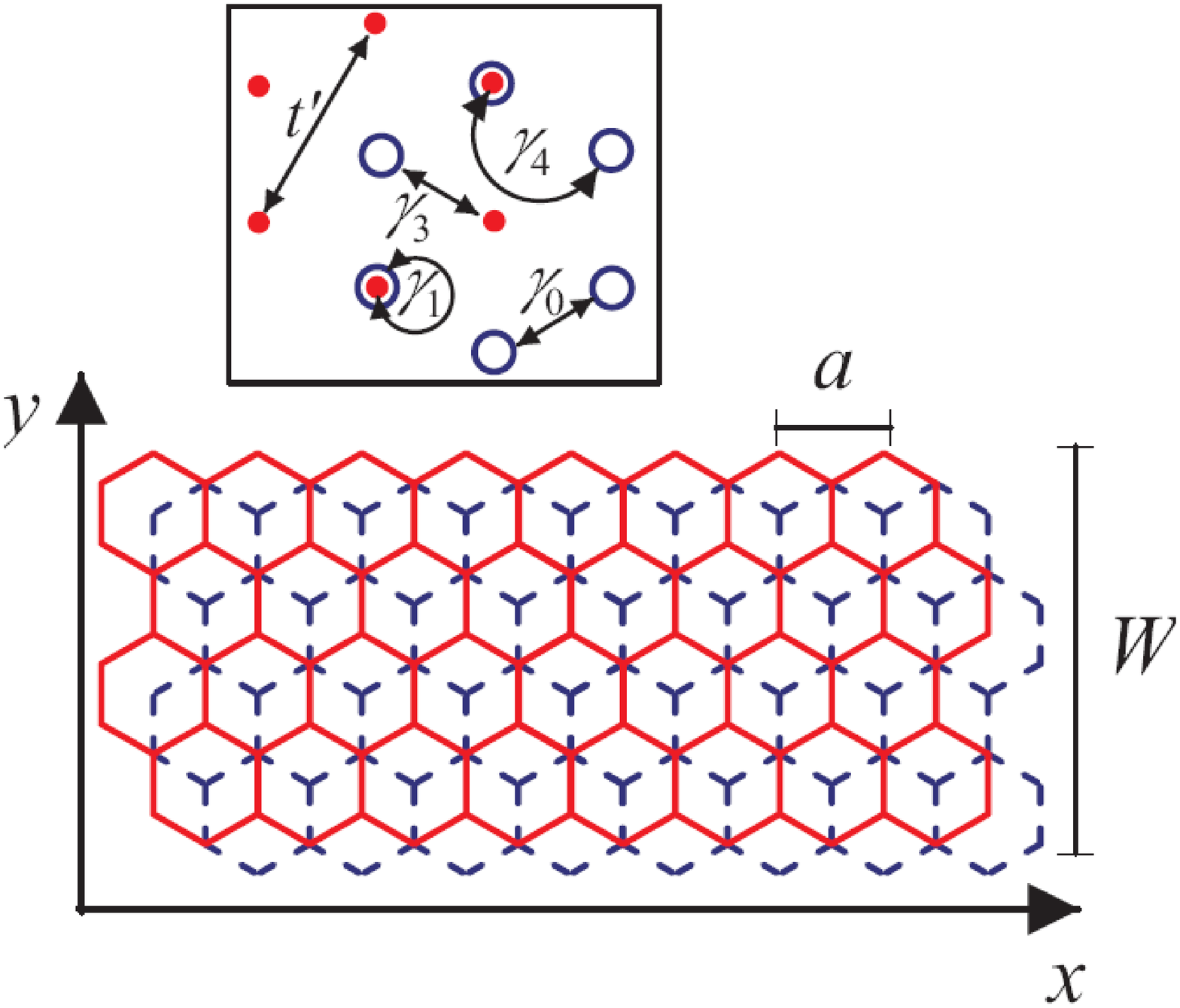}}
\includegraphics[width=\columnwidth]{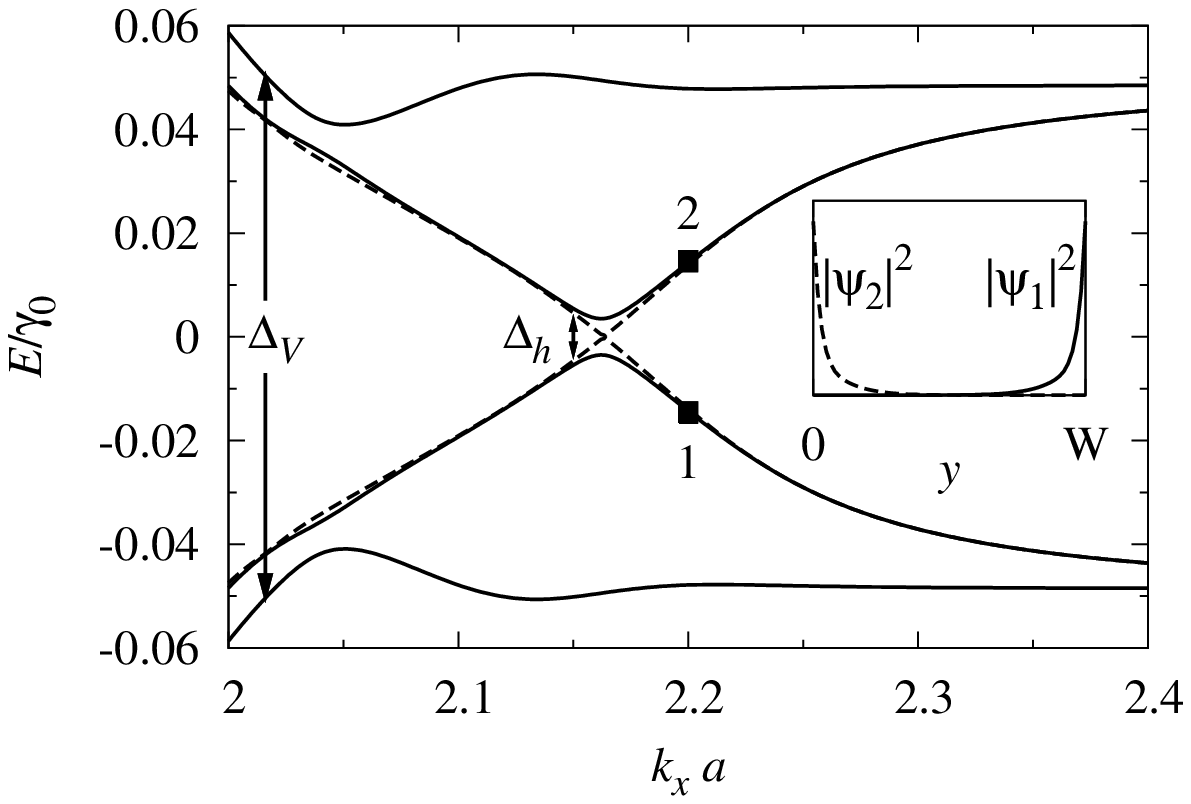}
\includegraphics[width=\columnwidth]{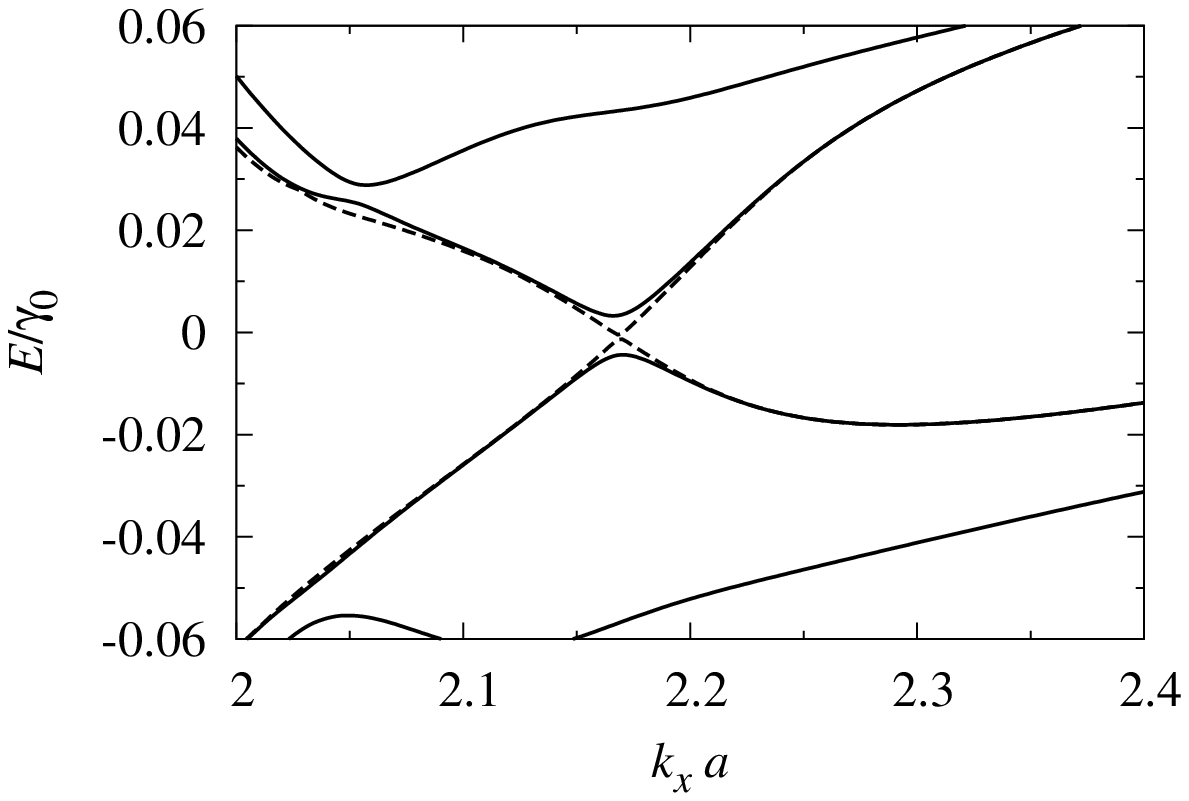}
\caption{\label{fig1} (Color online) Upper panel: Sketch of a
bilayer graphene nanoribbon of lattice constant $a$ and width $W$
(inset: definition of hopping terms). Middle panel: Dispersion
relation of edge states for a nanoribbon with
$W=32\times\sqrt{3}\, a$ and layer-symmetry breaking potential
$V=0.025\,\gamma_0$, focussing on one of the two K points
($k_0=2\pi/(3a)$). (Dashed lines: two innermost bands for
$W=40\times\sqrt{3}\, a$.) Here we use hopping matrix elements
$\gamma_1=\gamma_3=0.12\,\gamma_0$, but set $\gamma_4=t'=0$. The
inset shows the transverse charge distribution of two edge states
with energies indicated by the markers.  Lower panel: The same but
including hopping terms $\gamma_4=0.05\,\gamma_0$,
$t'=0.07\,\gamma_0$. }
\end{figure}

Previous works on edge states in gated bilayer ribbons ignore two
elements: (i) next-nearest-neighbor hopping, which is known to
strongly affect the bulk properties
\cite{malard:201401,li:037403,arxiv:0908.0672} and is of crucial
importance for edge states in monolayer ribbons,
\cite{sasaki:113110} and (ii) the effects of interactions, which
are known to strongly influence the properties of quasi-one
dimensional systems, including carbon-based systems
.\cite{Kane97,brey} In this paper, we point out that
next-nearest-neighbor hopping breaks the symmetry between the
edges, resulting into a slow edge and a fast edge characterized by
different values of the propagation velocity. We then explore the
consequences for the question whether many-body effects can help
to open a gap in these channels. Treating interactions in a
Hubbard model on the basis of bosonization and renormalization
group methods at $T=0$, we find that the edge channels display a
Mott transition at half filling. In presence of the higher-order
hopping terms, the associated charge gap $\Delta_\rho$ on the slow
edge can take on sizeable values, while on the fast edge the gap
is negligibly small. Remarkably, the gap depends sensitively on
the applied gate voltage $V$; the effects of interactions can
therefore be controlled externally. Furthermore, the dependence of
$\Delta_\rho$ on $V$ is non-monotonic, in striking contrast to the
behavior of the field-induced gap $\Delta_V$ in the bulk of the
system.

This paper is organized as follows. In Sec. II we present the
hamiltonian for the graphene bilayer ribbon including the next-nearest-neighbor
hopping and the hopping term between non-dimerised atoms in both layers. We then
describe the edge states for a zigzag ribbon of bilayer graphene and their
different localization properties along the transverse direction of the ribbon
by computing the Inverse Participation Ratio. In Sec. III we describe the effects
of considering electron-electron interactions in the spectrum for edge states.
We find that because the different properties of the two edges the gap opened
by the interaction is different for both edges. We also briefly discuss a
possible experimental verification of this different behavior. We summarize
our results in Sec. IV.

\section{Tight Binding model}

A graphene bilayer nanoribbons with zigzag edge termination is shown in Fig.\
\ref{fig1} (left panel). Focussing on the low-energy bands which participate
in electronic transport, such ribbons can be modelled by a tight-binding
hamiltonian
$H=\sum_{\sigma=\downarrow,\uparrow}\sum_{ij}\gamma_{ij}c_{i,\sigma}^\dagger
c_{j,\sigma}$, 
where each carbon atom hosts one spin-degenerate electronic
orbital with annihilation operator $c_{i\sigma}$. Recent experiments
specifically addressed the values of the parameters $\gamma_{ij}$ in bilayer
graphene. \cite{malard:201401,li:037403,arxiv:0908.0672} Nearest-neighbor
hopping in the same graphene layer is described by a hopping element
$-\gamma_0$, where $\gamma_0\simeq 3\,\mbox{eV}$. The layers are Bernal
stacked, with hopping element $\gamma_1\simeq 0.12 \gamma_0$ between
dimerised atoms that lie on top of each other. The symmetry between both
layers can be broken by top- or back-gates with a gate voltage $V_{g}$, which
induce a perpendicular electric field and give rise to an on-site potential
$\pm V$ on the two layers, which is obtained from $V_{g}$ by including the
screening in the layers. \cite{McCann06a} We assume that $V$ is uniform across
the system, even near the boundaries, which in reality will be enforced to a good
approximation by the proximity of the metallic gate close to the system. At the boundary,
we impose standard hard-wall boundary conditions,which amount to setting the
wavefunction to zero on lattice sites that lie outside the ribbon.

It is also conventional to include the direct coupling $\gamma_3$ between the
non-dimerised atoms in both layers. \cite{McCann06} Ignoring for the moment
other hopping terms, a typical dispersion relation of the edge states in a
gated ribbon is shown in Fig.\ \ref{fig1} (middle panel). Of the four bands
of edge states, two are massive, with a gap $\Delta_V\simeq 2|V|$ (the same
as for the bulk states), while the other two display a linear dispersion,
corresponding to two counter-propagating states which are well localized at
opposite edges (see inset). The same scenario is replicated at the other K
point in the graphene Brillouin zone, but with the edges interchanged.
Therefore, each edge supports two counter propagating state, one from each K
point. Notably, to this level of approximation, the propagation velocities on
both edges are the same.

\begin{figure}
\includegraphics[width=\columnwidth]{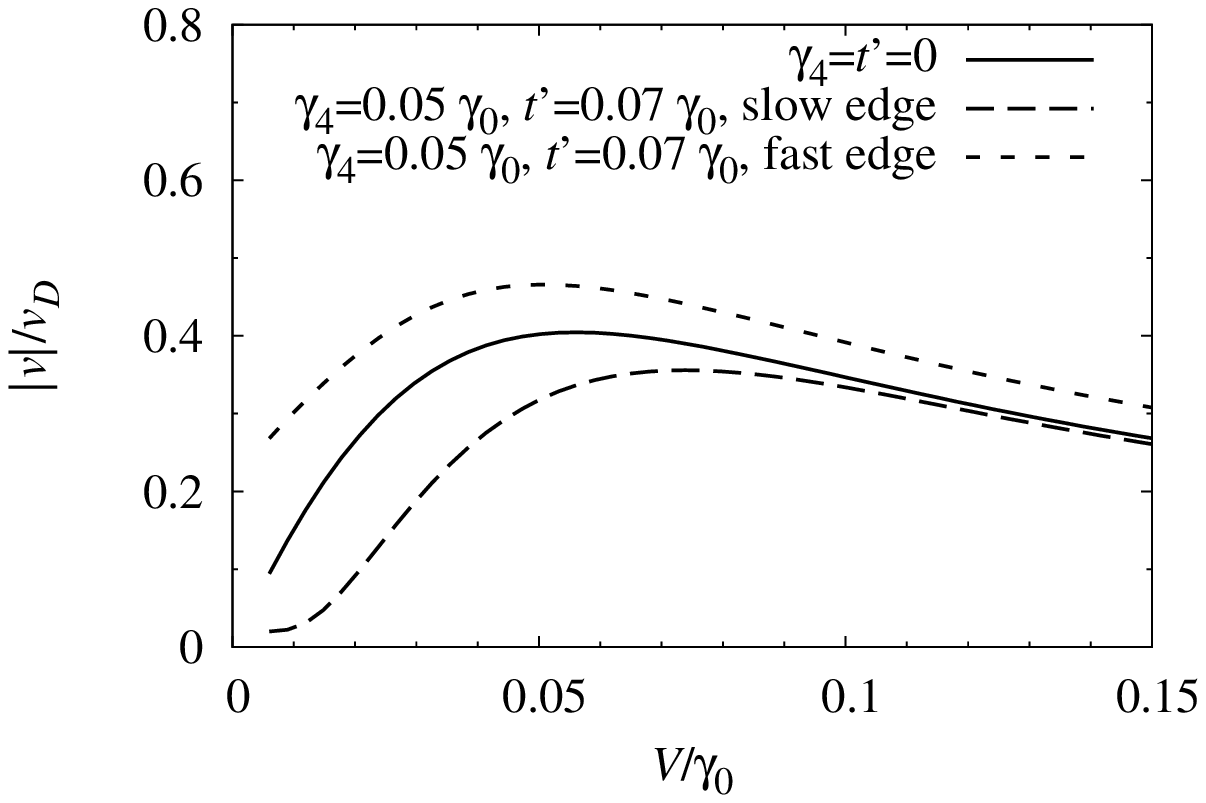}
\includegraphics[width=\columnwidth]{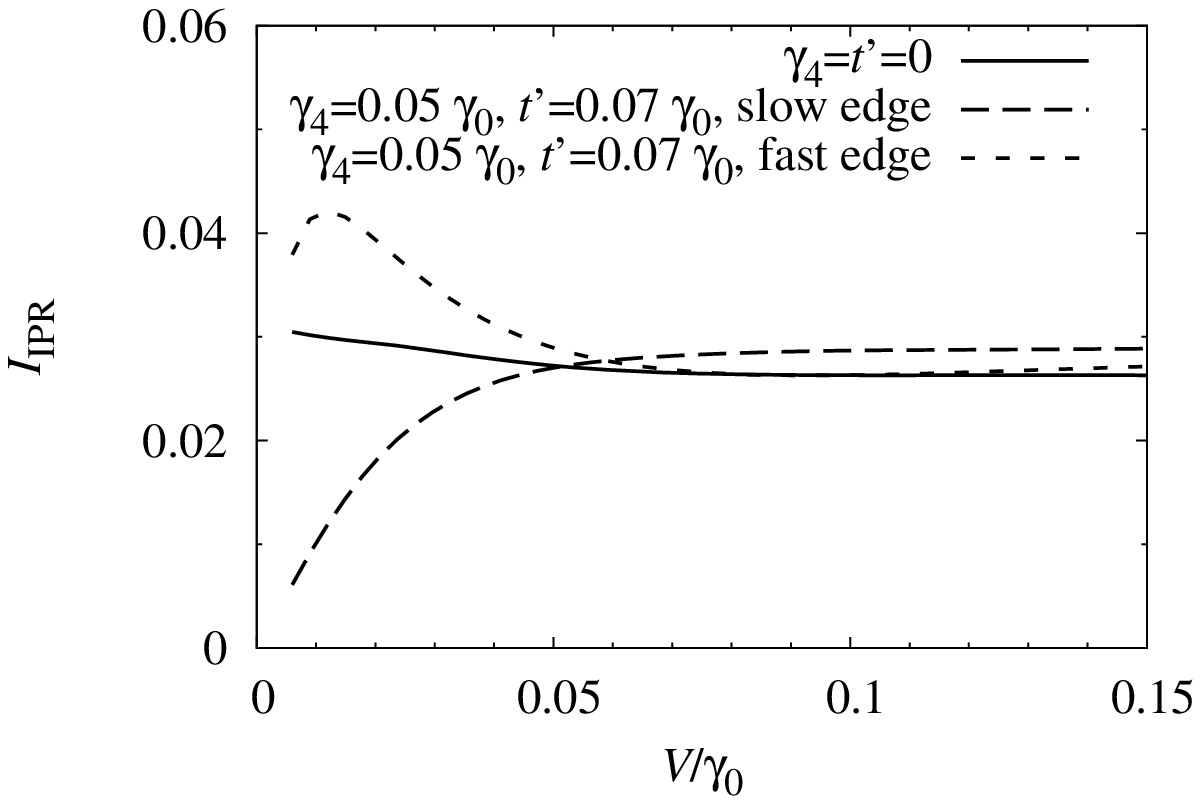}
\caption{\label{fig1a} Influence of $\gamma_4$ and $t'$ on the
edge-state velocities (top panel, $v_D=\sqrt{3}\gamma_0 a/2\hbar$) and 
localization on each isolated edge ($\Delta_{h}=0$)(bottom panel,
expressed in terms of the inverse participation ratio $I_{\rm
IPR}$). }
\end{figure}

The middle and lower panels of Fig.\ \ref{fig1} shows how the dispersion
changes when the two most prominent additional hopping elements
are taken into account: the interlayer hopping $\gamma_4\approx
0.05\,\gamma_0$ between adjacent dimerized and non-dimerized
carbon atoms, and the intralayer hopping $t'\approx
0.07\,\gamma_0$ between next-nearest neighbors. Both terms break
the particle-hole symmetry, which then discriminates the different
edges: the terminating atoms live on different layers and
therefore posses a different on-site potential. In the dispersion
relation, this takes the effect of an additional background
velocity, which increases the velocity at one edge (the 'fast
edge') while reducing the velocity at the other edge (the 'slow
edge'). The dependence of these velocities on the gate potential
is shown in the top panel of Fig.\ \ref{fig1a}. The additional
hoppings also affect the transverse localization of the edge
states. This is shown in the bottom panel in terms of the inverse
participation (IPR)
$I_{\rm IPR}=\sum_n|\psi_n|^{4}$,
which is larger the better localized a state is (here, the sum is
over the transverse direction of the nanoribbon).

At this point it is important to mention that there are two types of
zigzag edge termination in graphene bilayer,\cite{Sahu08} the so called
$\alpha$ and $\beta$ terminations. When the on site potential $V$ is zero
some qualitative differences in the bandstructure can be found in
both cases. However when $V_{g}$ leads to a nonvanishing $V$ in
both cases two of the edge bands become gapped (with a gap of the
order of $V$) and the two other bands cross each other around the
Fermi points as discussed before for the two types of edge terminations.
Also in both cases the asymmetry between the velocities at different
edges can be found, being the differences most prominent at the high
energy bands. Thus we will not consider any distinction between the two
terminations as long as we are only concerned in the physics at small
energies around the Fermi points.

At small energies, the hybridization of the edge states leads to
an avoided crossing, where the small residual gap $\Delta_h$
vanishes exponentially with increasing width of the ribbon. A
sizable hybridization gap $\Delta_h$ exists only for small values
of $W$ (narrow ribbons). For typical applied gate voltages and
widths, each edge of the nanoribbon therefore behaves like a
metallic one-dimensional system with two counter-propagating
states, where each state is associated to one of the K points.
In this scenario of low dimensionality, one should expect that 
interactions play an important role. Generally, however, interactions
are felt most strongly when particles propagate slowly and are well 
confined. Thus, the concrete manifestations should depend on the distinct
properties of the fast and slow edge states described above. This
is what we will explore in the remainder of this paper.

\section{Many body effects}

Within the tight-binding description, interactions can be
incorporated via a Hubbard term
$U\sum_{i}n_{i,\uparrow}n_{i,\downarrow}$, where
$n_{i,\sigma}=c_{i,\sigma}^\dagger c_{i,\sigma}$ is the occupation
operator on lattice site $i$ with electrons of spin
$\sigma=\downarrow,\uparrow$. Assuming $\Delta_h=0$,  each edge
can be treated separately via a hamiltonian of non-chiral bosons:
\begin{equation}
H_{0}=\sum_{\mu=\rho,\sigma}\int{dx}\frac{v_{\mu}K_{\mu}}{2}\left(\frac{\partial
\phi_{\mu}}{\partial x}\right)^{2}
+\frac{v_{\mu}}{2K_{\mu}}\left(\frac{\partial
\theta_{\mu}}{\partial x}\right)^{2},\label{nonintbosh}
\end{equation}
where $\mu$ labels the charge ($\rho$) and spin ($\sigma$) sectors, while
$v_{\mu}$ and $K_{\mu}$ are the velocity and the Luttinger parameters,
respectively. Interaction events that do not lead to backscattering
renormalize these parameters in different ways, which
yields the usual separation between the spin and charge degrees of freedom
\cite{Giamarchi}:
\begin{equation}
v_{\rho,\sigma}=\sqrt{\left(v\pm\frac{g_{f}}{2\pi\hbar}\right)^{2}-\left(\frac{g_{b}}{2\pi\hbar}\right)^{2}},\label{vcharge}
\end{equation}
\begin{equation}
K_{\rho,\sigma}=\sqrt{\frac{\hbar
v\pm\frac{g_{f}}{2\pi}\mp\frac{g_{b}}{2\pi}}{\hbar
v\pm\frac{g_{f}}{2\pi}\pm\frac{g_{b}}{2\pi}}},\label{Kcharge}
\end{equation}
where the upper sign applies to $\rho$, while the lower sign
applies to $\sigma$.

Most of the interesting physics arises from backward scattering,
which leads to a nonlinear Sine-Gordon term in the spin sector,
\begin{equation}
H_{\rm back}=\frac{g_{b}}{2\pi^{2}a^{2}}\int{dx}\cos
(\sqrt{8}\phi_{\sigma}).\label{sinegordonspin}
\end{equation}

At half filling, which we need to consider to assess the size of
any interaction-induced gap, this is accompanied by umklapp
processes,
which contribute an analogous term to the charge sector:
\begin{equation}
H_{\rm umklapp}=\frac{g_{b}}{2\pi^{2}a^{2}}\int{dx}\cos
(\sqrt{8}\phi_{\rho}).\label{sinegordoncharge}
\end{equation}

All these terms are controlled by the forward and backward
coupling constants $g_{b,f}$, which are obtained
 by writing the Hubbard interaction term
$U n_{i,\uparrow} n_{i,\downarrow}$ in the basis
\begin{equation}
\Psi(x,y)\sim
\varphi_{-}(y)e^{-ik_0x}L(x)+\varphi_{+}(y)e^{ik_0x}R(x)\label{1poperator}
\end{equation}
of the diagonalized one-particle hamiltonian around each K point
($k_0=2\pi/(3a)$):
\begin{subequations}
\begin{eqnarray}
g_{f}&=&U a^2\int{dy}|\varphi_{\pm}(y)|^{2}|\varphi_{\pm
}(y)|^{2},\label{forwardg}
\\
g_{b}&=&U a^2
\int{dy}|\varphi_{\pm}(y)|^{2}|\varphi_{\mp}(y)|^{2}.\label{backwardg}
\end{eqnarray}
\end{subequations}
The left-right symmetry of the system results in the additional constraint $
g_f= g_b =g=U a I_{\rm IPR}$,  which involves the IPR of the transverse
wavefunction. As seen above,  as soon as $\gamma_4$ and $t'$ are taken into
account the IPR strongly depends on the layer-symmetry breaking potential
$V$. Thus, the effective strength of interactions can be controlled by the
applied gate voltage.


Since the terms (\ref{sinegordonspin}) and (\ref{sinegordoncharge}) make the
hamiltonian not exactly solvable, we assess their consequences using standard
renormalization group (RG) arguments. \cite{Giamarchi,Tsvelik} In first order
in $g_{b}$, the RG equations for $y_\mu \equiv \frac{g_{b}}{\hbar\pi v_\mu}$
and $K_\mu$  are
\begin{eqnarray}
\frac{d K_\mu}{d\textit{l}}=-\frac{y_\mu^{2}}{2},\label{RGeqs}
\quad \frac{d
y_\mu}{d\textit{l}}=\left(2-2 K_\mu\right)y_\mu,
\end{eqnarray}
where $l$ is a logarithmic renormalization scale. The scaling
behavior of $y_\mu$ depends on the value of $K_\mu$. For
$K_\mu>1$, $y_\mu$ decreases when the scaling parameter $\emph{l}$
increases, which renders the interaction term irrelevant. For
$K_\mu<1$, however, the interactions are relevant; the
intermediate case $K_\mu=1$ represents a quantum critical point.

\begin{figure}
\includegraphics[width=\columnwidth]{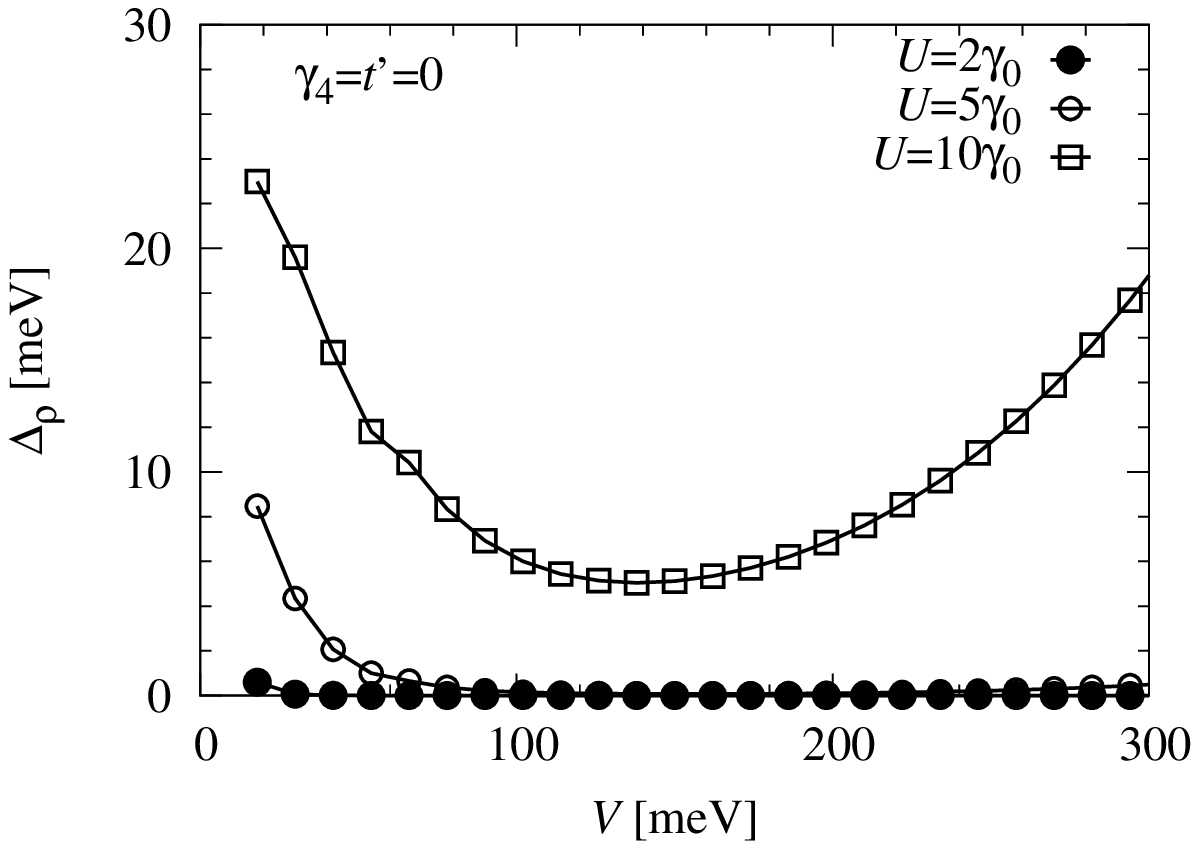}
\includegraphics[width=\columnwidth]{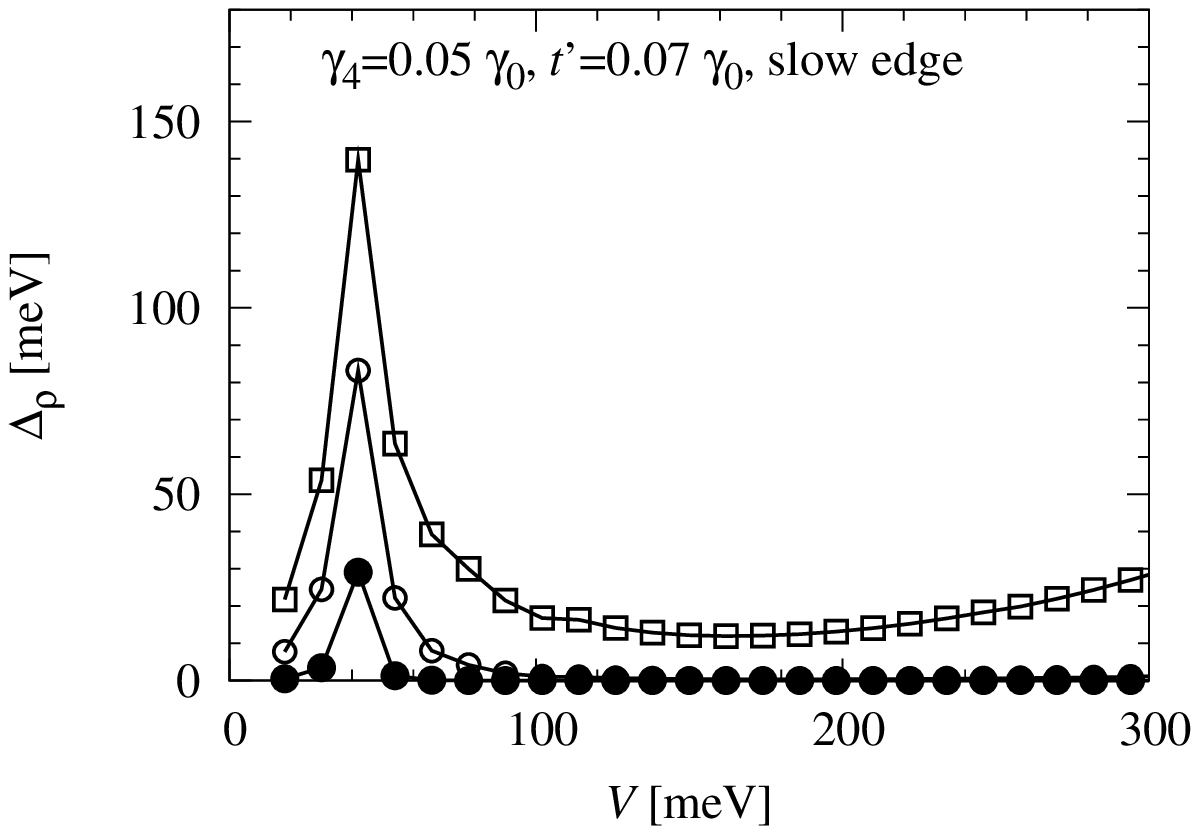}
\includegraphics[width=\columnwidth]{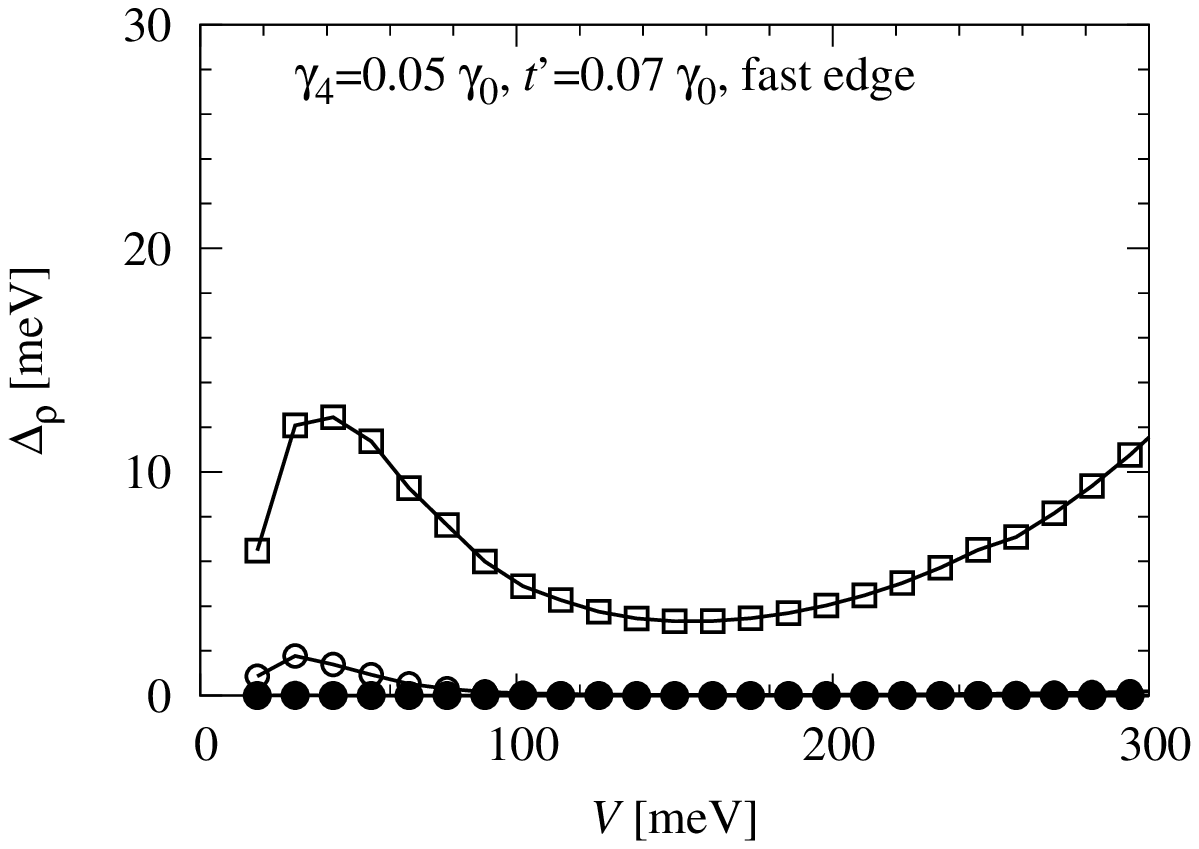}
\caption{\label{fig3} Charge gap $\Delta_{\rho}$ as a function of
$V$, for selected values of $U$, assuming $\gamma_4=t'=0$ (upper
panel), as well as $\gamma_4=0.05\,\gamma_0$, $t'=0.07\,\gamma_0$
(middle panel: slow edge; lower panel: fast edge). Note the
different scales of the vertical axis.}
\end{figure}

Because of the constraint $g_{f}= g_{b}= g$, the Luttinger
parameter (\ref{Kcharge}) can be written as
\begin{equation}
K_{\rho}=\sqrt{\frac{1}{1+\frac{g}{\pi \hbar v}}}, \quad
K_{\sigma}=\sqrt{\frac{1}{1-\frac{g}{\pi \hbar v}}}.
\end{equation}
For the  case $g>0$ (repulsive interactions), to which we restrict our
attention, $K_{\sigma}>1$ and the RG equations (\ref{RGeqs}) imply that the
spin sector will flow towards the noninteracting theory, with renormalized
$K_{\sigma}^{*}=1$ due to the emerging spin-rotation invariance. On the other
hand, $K_{\rho}$ is smaller than one, and therefore the system may acquire a
gap in the charge sector. For a more quantitative analysis, we introduce the
new variable $x_\rho=2K_\rho-2$, so that Eq.\ (\ref{RGeqs}) takes the form
\begin{eqnarray}
\frac{d x_\rho}{d\textit{l}}=-y_\rho^{2},\quad \frac{d
y_\rho}{d\textit{l}}=-x_\rho y_\rho. \label{RGeqs2}
\end{eqnarray}
In these new variables, the RG equations  possess a first integral
$A^{2}=x_\rho^{2}-y_\rho^{2}=x_{\rho,0}^{2}-y_{\rho,0}^{2}$, whose
value remains constant along the  trajectories of the flow, and
therefore can be calculated using the bare values $x_{\rho,0}$ and
$y_{\rho,0}$. Most importantly, none of the flow lines cross the
line $x_\rho=y_\rho$, where $A=0$. This line constitutes a
separatrix between two regimes. For $y_\rho<x_\rho$, $y_\rho$
scales to zero and $x_\rho$ approaches a definite renormalized
value $x_\rho^{*}$. Physically, this sector of the system is then
well described by the noninteracting hamiltonian
(\ref{nonintbosh}) with renormalized $K_\rho^{*}$. For
$y_\rho>x_\rho$, however, $y_\rho\to \infty$ and $x_\rho\to
-\infty$ flow towards strong coupling, where the RG equations are
no longer valid. The charge sector then acquires a gap
$\Delta_\rho$.

For bare value $|x_{\rho,0}|/y_{\rho,0}\ll 1$, this gap can be estimated
using the selfconsistent harmonic approximation, \cite{Tsvelik} which
exploits that the gap scales with $l$ as $\Delta_\rho=\Delta_{0} e^{l}$, with
$\Delta_{0}\sim v_{\rho}\Lambda$. This leads to $ \Delta_\rho=\hbar
v_\rho\Lambda\left(\frac{4K_\rho
y_\rho}{(a\Lambda)^{2}}\right)^{\frac{1}{2-2K_\rho}}$, where we equate the
real-space $a$ cutoff with the lattice constant. The quantity $\Lambda$ is an
ultraviolet cutoff in the momentum space. Since the edge dispersion relation
is linear only for energies less than $\Delta_V\approx 2V$, $\Lambda \sim
\frac{2V}{\hbar v_{\rho}}$, which delivers
\begin{equation}
\Delta_{\rho}=2V\left(\frac{K\hbar v_{\rho} g}{\pi a^{2}
V^{2}}\right)^{\frac{1}{2-2K_{\rho}}}.\label{gap2}
\end{equation}
However, this estimate is only valid when $\left|x_{\rho,0}\right|\ll
y_{\rho,0}\ll 1$. For $\left|x_{\rho,0}\right|\lesssim y_{\rho,0}\ll 1$, the
gap must be obtained by terminating the flow at $y_\rho\sim 1$, where the
first-order expansion breaks down, giving
\begin{equation}
\Delta_{\rho}=2V\sqrt{1+y_{\rho}}e^{-1/y_{\rho}}=2V (v_{\rho}/v)e^{-\frac{\pi \hbar
v}{g}}.\label{alternativegap}
\end{equation}

In order to determine which of these regimes applies to the bilayer graphene
edge states, we have computed the ratio $x/y$ for some
values of $V$ and $U$. Consistently, $|x_{\rho,0}|/y_{\rho,0}\lesssim 1$,
which means that the gap should be obtained from Eq.\ (\ref{alternativegap}),
where $v$ and $g=Ua I_{\rm IPR}$ follow from the results of Fig.\
\ref{fig1a}.

The size of the resulting gap for different scenarios is shown in
Fig.\ \ref{fig3}. In all cases, the dependence of $\Delta_\rho$ on
$V$ is non-monotonic. Ignoring next-nearest-neighbor hopping (left
panel), the gap at both edges is identical, but takes a sizeable
value only for unrealistically large values of the Hubbard
parameter. When $\gamma_4$ and $t'$ are taken into account, the
gap at the fast edge (right panel) is further suppressed. However,
the gap at the slow edge (middle panel) is dramatically increased,
in particular in the region where the propagation velocity becomes
small.

\section{Conclusions}

We have studied the effect of next-nearest-neighbor
hopping and interactions on the edge states in gated bilayer
graphene nanoribbons with zigzag termination. The additional
hopping results in the formation of a slow and a fast edge, where
electrons propagate with different velocities. The small velocity
at the slow edge assists the formation of an interaction-induced
gap $\Delta_\rho$. Whether a sizeable gap can be achieved at
realistic interaction strengths would be best decided by an
experiment which exploits that umklapp processes are suppressed
when the temperature is raised. The physics behind any
experimentally observed gap can therefore be probed via the
temperature dependence of the conductivity (measured at the edge
via side contacts), which for $T\gg\Delta_\rho$ should scale as
$\sigma\propto T^{3-4K_{\rho}}$. \cite{Giamarchi91} The experimental
set up can be a four-terminal geometry which probes each edge separately 
($\Delta_{h}=0$ and thus the two edges are decoupled for a sufficiently 
wide nanoribbon). Because each edge state possesses different a 
different Fermi velocity $K_{\rho}$ is
thus different as it is discussed in the text allowing to
measure independently each conductivity. This algebraic
dependence is distinct from the exponential behavior of the
conductivity when the gap originates in the single-particle
spectrum (such as the hybridization gap $\Delta_h$ due to the
finite width of the ribbon). The simultaneous presence of the
hybridization gap and interactions could be modelled by a
bosonized interaction term $H_{\rm gap}=-\frac{\Delta_h}{\pi^{2}
\alpha^{2}}\sin(\sqrt{8}\phi_{\rho})\cos(\sqrt{8}\phi_{\sigma}) $
,\cite{Varma85,Balents96} which couples the spin- and charge
sectors. Moreover, in this situation, right- and left-moving
states around the same K point become coupled because of their
finite overlap. The fact that these states propagate with
different velocities adds interesting complications, which,
however, go beyond the scope of the present work.

\section{Adknowledgements}

We gratefully acknowledge discussions with F. Guinea, M. A. H. Vozmediano and
G. Le{\'o}n Suros and support by the EC via Grant No. MEXT-CT-2005-023778.


\begin{thebibliography}{0}
\expandafter\ifx\csname natexlab\endcsname\relax\def\natexlab#1{#1}\fi
\expandafter\ifx\csname bibnamefont\endcsname\relax
  \def\bibnamefont#1{#1}\fi
\expandafter\ifx\csname bibfnamefont\endcsname\relax
  \def\bibfnamefont#1{#1}\fi
\expandafter\ifx\csname citenamefont\endcsname\relax
  \def\citenamefont#1{#1}\fi
\expandafter\ifx\csname url\endcsname\relax
  \def\url#1{\texttt{#1}}\fi
\expandafter\ifx\csname urlprefix\endcsname\relax\def\urlprefix{URL }\fi
\providecommand{\bibinfo}[2]{#2}
\providecommand{\eprint}[2][]{\url{#2}}

\end{thebibliography}


\begin{thebibliography}{99}

\bibitem{TaisukeOhta08182006} T. Ohta et al.,
  Science \textbf{313}, 951 (2006).

\bibitem{McCann06} E. McCann and  V.~I. Fal'ko,
  Phys. Rev. Lett. \textbf{96}, 086805 (2006).

\bibitem{McCann06a} E. McCann,
  Phys. Rev. B \textbf{74}, 161403(R) (2006).

\bibitem{Castro07} E.~V. Castro et al.,
  Phys. Rev. Lett. \textbf{99}, 216802 (2007).

\bibitem{Min08}
H. Min, G. Borghi, M. Polini, and A. H. MacDonald, Phys. Rev. B
\textbf{77}, 041407(R) (2008).


\bibitem{Jiao09} L. Jiao et al.,
  Nature \textbf{458}, 877 (2009).

\bibitem{Campos09} L.~C. Campos et al.,
  Nano Letters \textbf{9}, 2600 (2009).

\bibitem{Han07} M.~Y. Han et al.,
  Phys. Rev. Lett. \textbf{98}, 206805 (2007).

\bibitem{Castro08} E.~V. Castro et al.,
  Phys. Rev. Lett. \textbf{100}, 026802 (2008).

\bibitem{Rhim08} J. Rhim and  K. Moon,
  J. Phys.: Condens. Matter \textbf{20}, 365202 (2008).

\bibitem{Sahu08} B. Sahu, H. Min, A.~H. MacDonald, and S. Banerjee,
  Phys. Rev. B \textbf{78}, 045404 (2008).

\bibitem{Lima09} M.~P. Lima, A. Fazzio, and A.~J.~R. da~Silva,
  Phys. Rev. B \textbf{79}, 153401 (2009).

\bibitem{Rycerz07} A. Rycerz, J. Tworzyd{\l}o, and C.~W.~J. Beenakker,
  Nat. Phys. \textbf{3}, 172 (2007).

\bibitem{malard:201401} L.~M. Malard et al.,
Phys. Rev. B \textbf{76}, 201401  (2007).

\bibitem{li:037403} Z.~Q. Li et al.,
  Phys. Rev. Lett. \textbf{102}, 037403 (2009).

\bibitem{arxiv:0908.0672} A.~B. Kuzmenko et al.,
  arXiv:0908.0672.

\bibitem{sasaki:113110} K. Sasaki, S. Murakami, and R. Saito,
  App. Phys. Lett. \textbf{88}, (2006).

\bibitem{Kane97} C. Kane, L. Balents, and P.~A. Fisher,
  Phys. Rev. Lett. \textbf{79}, 5086 (1997).


\bibitem{brey} L. Brey and H. A. Fertig,
  Phys. Rev. B \textbf{75}, 125434 (2007).

\bibitem{Giamarchi} T. Giamarchi, \emph{Quantum Physics in One Dimension}
  (Oxford Science Publications, Oxford, 2003).


\bibitem{Tsvelik} A.~O. Gogolin, A.~A. Nersesyan, and A.~M. Tsvelik,
  \emph{Bosonization and Strongly Correlated Systems}
  (Cambridge University Press, Cambridge, 1998).

\bibitem{Giamarchi91} T. Giamarchi,
  Phys. Rev. B \textbf{44}, 2905 (1991).

\bibitem{Varma85} C.~M. Varma and A. Zawadowski,
  Phys. Rev. B \textbf{32}, 7399 (1985).

\bibitem{Balents96} L. Balents and M.~P.~A. Fisher,
  Phys. Rev. B \textbf{53}, 12133 (1996).

\end{thebibliography}
\end{document}